\documentclass{article}
\usepackage[top=30truemm,bottom=30truemm,left=25truemm,right=25truemm]{geometry}
\usepackage{mathrsfs}
\usepackage{cite}
\usepackage{authblk}
\usepackage[dvips]{graphicx}
\usepackage{bm,amssymb}
\usepackage{amsmath}
\usepackage{txfonts}
\usepackage{booktabs}
\usepackage{cases}
\usepackage{color}
\usepackage[normalem]{ulem}  % \sout{old text} for strikeout

\renewcommand\sout{\bgroup \color{red} \ULdepth=-.5ex \ULset}
\newcommand{\Slash}[1]{{\ooalign{\hfil/\hfil\crcr$#1$}}}

\def\pbar{\overline{\psi}}

\def\Gs2{\Gamma^{(2)}_{k,\sigma}}
\def\Gp2{\Gamma^{(2)}_{k,\pi}}

\def\Jab{J_{k,\alpha\beta}}
\def\Jss{J_{k,\sigma\sigma}}
\def\Jpp{J_{k,\pi\pi}}
\def\Jsp{J_{k,\sigma\pi}}
\def\Jps{J_{k,\pi\sigma}}

\def\Is{I^{(2)}_{k,\sigma}}
\def\Ip{I^{(2)}_{k,\pi}}

\def\Jsqq{J^{(\sigma)}_{k,\bar{\psi}\psi}}
\def\Jpqq{J^{(\pi)}_{k,\pbar\psi}}

\def\Dk{\partial_{k}}

\def\RB{R^{B}_{k}}
\def\RF{R^{F}_{k}}

\def\Ea{E_{\alpha}}
\def\Eb{E_{\beta}}

\def\Ebt{\tilde{E}_{\beta}}

\def\ip0{\mathrm{i}p_{0}}
\def\p02{p_{0}^{2}}

\def\q2{\vec{q}^2}

\newcommand{\MeV}{\mathrm{MeV}}
\newcommand{\pab}{
\ifmmode p
\else $p$
\fi
}
\newcommand{\pfour}{
\ifmmode P
\else $P$
\fi
}
\newcommand{\qfour}{
\ifmmode Q
\else $Q$
\fi
}

\begin{document}
\title{Spectral functions in functional renormalization group approach\\
\Large{--
%application to the
analysis of the collective soft modes at the QCD critical point --}
}
\author{Takeru Yokota$^{1}$, Teiji Kunihiro$^{1}$ and Kenji Morita$^{2}$}
\affil{$^1$Department of Physics, Faculty of Science, Kyoto University, Kyoto 606-8502, Japan\\
 $^2$Yukawa Institute for Theoretical Physics, Kyoto University, Kyoto 606-8502, Japan}
\maketitle

\begin{abstract}
We first review the method to calculate the spectral functions in the functional renormalization group
(FRG) approach, which has been recently developed.
We also provide the numerical stability conditions given by the present authors
for a generic nonlinear evolution equation
that are necessary for obtaining the accurate effective potential from the flow equation in the FRG.
As an interesting example, we report the recent calculation of  the spectral functions of the mesonic and particle-hole
excitations using a chiral effective model of Quantum Chromodynamics (QCD);
we extract the dispersion relations from them and
try to reveal the nature of the soft modes at the QCD critical point (CP) where the phase transition is second order.
Our result shows that a clear development and the softening of
the phonon mode in the space-like region  as the system approaches the CP;
furthermore  it turns out that
the sigma mesonic mode once in the time-like region gets to merge
 with the phonon mode in the close vicinity of the CP, implying a novel
possibility about the nature of the soft mode of the QCD CP.
\end{abstract}

\section{Introduction}
\label{Intro}
The functional renormalization group (FRG)
\cite{Wetterich:1992yh,Wegner:1972ih,Wilson:1974,Polchinski:1983gv}
is a non-perturbative method of the field theory
which enables us to investigate strongly correlated systems
 with incorporation of fluctuation effects beyond the mean-field theory.
The FRG has been applied to a wide range of fields \cite{Berges:2000ew,Pawlowski:2005xe,Gies:2006wv,Jungnickel:1995fp,
Braun:2003ii,Schaefer:2005,Schaefer:2006ds,Stokic:2010, Nakano:2010, Aoki:2014}.
The FRG has proved powerful to study the equilibrium state of
many-body systems or the nonperturbative vacuum of a quantum field theory.
Elucidating dynamical properties of a physical system including possible emergence of collective excitations
is important for fully understanding the physical properties of the system.
It is to be noted that a phase change of the system can manifest itself as those in
the properties of elementary excitations or more generally in the spectral functions
in specific channels.
Thus it is notable that
 the calculation of spectral functions has become possible in the framework
of FRG\cite{Kamikado:2013sia,Tripolt:2013jra,Tripolt:2014wra,Yokota:2016tip}, and hence
one can extract
the characteristics of system such as the possible development of collective
modes and dispersion relations of modes\cite{Yokota:2016tip}.

Usually, the FRG applied to finite-temperature ($T$) systems is formulated in the imaginary-time formalism.
A real-time analysis is, however,  needed for
extracting the spectral functions for excitation modes, which are essentially
given as the imaginary part of
the retarded Green's function, and
an analytic continuation of two-point functions from imaginary Matsubara frequencies
to real frequencies is made to have the real-time two-point Green's
functions.
It is, however, not a simple task and can be even quite intricate to
perform an analytic continuation to get the spectral function in the nonperturbative method
\cite{Jarrell:1996,Asakawa:2000tr,Vidberg:1977,Dudal:2013yva}.
In the recent development\cite{Kamikado:2013sia,Tripolt:2013jra,Tripolt:2014wra,Yokota:2016tip},
an unambiguous way of the analytic
continuation has been proposed in the imaginary-time formalism, which
has turned out to lead to reasonable
results for the  spectral functions in the O(4) model in vacuum\cite{Kamikado:2013sia},
in  the quark-meson model at finite $T$ and chemical potential {$\mu$} \cite{Tripolt:2013jra,Tripolt:2014wra}.

The method has been adopted with some adaptation to
elucidate the nature of the soft modes of the critical point of
 Quantum Chromodynamics (QCD) at finite $T$ and $\mu$ \cite{Yokota:2016tip}.
The QCD phase diagram, i.e. the phase diagram for the system of quarks and gluons, is expected to have a rich structure
and its clarification is one of the hot topics
in the high-energy and nuclear physics \cite{Fukushima:2010bq}.
The QCD Lagrangian reads
\[
{\cal L}_{QCD}=\bar{\psi}(i\gamma_{\mu}D^{\mu}\,-\,\mbox{\boldmath $m$})\psi\,-\, \frac{1}{4}F_{\mu\nu}^a F^{\mu\nu a}
\]
where
$D_{\mu}=\partial_{\mu}-ig\,t^aA_{\mu}^a$ is the covariant derivative with $A_{\mu}^a$ being the gluon field with a color $a$ $(a=1,2,\cdots, N_c^2-1)$. $\gamma_\mu$ denote Dirac matrices.
Here
$\mbox{\boldmath $m$}={\rm diag}\,(m_u,\,m_d,\, \dots)$
 denotes the current quark mass matrix.
One of the key concepts of QCD is the chiral symmetry, which
is the invariance under the following independent two
transformations (collectively called chiral transformation):
\begin{equation}
\psi_{L}\rightarrow \exp\left(i\theta_{L}^{i}T^{i}/2\right) \psi_{L},\qquad
\psi_{R}\rightarrow \exp\left(i\theta_{R}^{i}T^{i}/2\right) \psi_{R},
%\notag
\label{eq:def-chiral-trans-form}
\end{equation}
where the $\psi_{L}$ and $\psi_{R}$ are left- and right-handed quark fields, respectively,
defined as $\psi_{L}=(1-\gamma_{5})\psi/2$ and $\psi_{R}=(1+\gamma_{5})\psi/2$ for quark field $\psi$
with $\gamma_{5}=i\gamma_{0}\gamma_{1}\gamma_{2}\gamma_{3}$.
The matrix $\gamma_5$ has the eigenvalues $\pm 1$, which are called the chirality (handedness);
$\psi_{L(R)}$ has the chirality $-1 \,(+1)$, and hence the name of chiral symmetry.
If we consider the $N_{f}$ flavor case where the quark field $\psi=\psi^{f j}$ has $N_f$
components as well as the color degrees of freedom ($j=1, 2,\dots, N_c$),\,
$\lbrace T^{i}\rbrace$ ($i=0,\cdots ,N_{f}(N_{f}-1)$)
are the generators for U($N_{f}$) transformation for flavor index. $\theta_{L}^{i}$
and $\theta_{R}^{i}$ are real global parameters.
Thus the transformations defined in (\ref{eq:def-chiral-trans-form}) form a group
$\mathrm{U(N_f)_L}\otimes \mathrm{U(N_f)_R}$
where the subscript $L\,(R)$ is attached to discriminate the vector space to
be transformed.
The chiral group includes a subgroup $\mathrm{U_{V}}(N_f)\simeq \mathrm{U_{V}}(1)\otimes\mathrm{SU_{V}}(N_f)$
which is realized
when the constraint on the group parameters $\theta_{L}^{i}=\theta_{R}^{i}\equiv \theta^i$\,
 ($i=0,\,1,\,2,\,\dots ,\, N_f(N_f-1)$)
 is imposed: In fact, this transformation is simply represented in terms of the quark field $\psi$ as
$\psi\,\to\,\exp\left(i\theta^{i}T^{i}/2\right) \psi.$.
We can also define the transformation with another constraint
$\theta_{L}^{i}=-\theta_{R}^{i}$, which is called
$\mathrm{U_{A}}(N_f)\simeq \mathrm{U_{A}}(1)\otimes \mathrm{SU_{A}}(N_f)$
transformation but does not form any group.
If the current quark masses are ignored, the chiral symmetry becomes an exact symmetry of the (classical)
QCD Lagrangian
because the vector current is written in terms of left- and right-handed fields separately;
$\bar{\psi}\gamma_{\mu}\psi=\bar{\psi}_L\gamma_{\mu}\psi_L,+\,\bar{\psi}_R\gamma_{\mu}\psi_R $, in contrast
to the Dirac mass term or scalar density $\bar{\psi}\psi=\bar{\psi}_R\psi_L\,+\,\bar{\psi}_L\psi_R$.
One also readily sees that chiral symmetry is explicitly broken
due to the current quark mass term $\bar{\psi}\mbox{\boldmath $m$}\psi$, although the neglect of this term is
a good approximation  in the low-energy regime for the lightest three flavors.
An important remark  is in order here:
It turns out that the $\mathrm{U_{A}}(1)$ symmetry is broken due to a quantum effect of QCD
called axial or $\mathrm{U_{A}}(1)$ anomaly\cite{'tHooft:1976up}.
Thus QCD in the quantum level has
a $\mathrm{U_{V}}(1)\otimes\mathrm{SU_{V}}(N_f)\otimes\mathrm{SU_{V}}(N_f)$ symmetry for
the massless $N_f$ flavors.

As Nambu first advocated\cite{Nambu:1960xd,Nambu:1961},
the (approximate) chiral symmetry is spontaneously broken in the
real world, and the pions are the massless bosons associated with the symmetry breaking
(now called the Nambu-Goldstone bosons) with a small mass $m_{\pi}$
acquired due to the small explicit breaking of the chiral symmetry for the two flavors.
Indeed some low-energy theorem (Gell-Mann-Oakes-Renner relation\cite{GellMann:1968rz})
tells us that the following formula holds;
\[
f_{\pi}^2m_{\pi}^2=-\frac{1}{2} (m_u+m_d) \langle \bar{\psi}\psi\rangle,
\]
where
$f_{\pi}\simeq 93$\,MeV is the pion decay constant with $\langle \bar{\psi}\psi \rangle\equiv \sigma_0$ denoting the
vacuum expectation value of the (isoscalar) scalar density $\bar{\psi}\psi =\bar{u}u+\bar{d}d$.
The existence of the finite scalar condensate (also called the chiral condensate) implies
that chiral symmetry is spontaneously broken and the chiral condensate $\sigma_0$
can be regarded as  an order parameter of the chiral transition of the QCD vacuum.

Apart from the chiral symmetry breaking,
elementary excitations  on top of the nonperturbative QCD vacuum are all color-singlet
and called hadrons, which are the manifestation of  the color confinement;
colored quarks and gluons do not exist as asymptotic states.
Thus at low-temperature and low-density regime, we have the confined phase with the chiral symmetry being
spontaneously broken, which we call the hadronic phase.
As in the usual many-body systems with a spontaneous symmetry breaking,
the  chiral symmetry is to be restored at high temperature and/or density where
colors may be also liberated:
Such a state of the matter is
called a quark--gluon plasma (QGP).

Effective chiral models, i.e. models focusing on the chiral symmetry,
has been utilized for an analysis of phase transitions in QCD.
One of such models is the celebrated Nambu--Jona-Lasinio (NJL) model \cite{Nambu:1961,Hatsuda:1994pi}.
In the case of $N_{f}=2$, the model Lagrangian reads
\begin{equation}
\mathscr{L}=\overline{\psi}\left(
i\Slash{\partial}-
\mbox{\boldmath $m$} \right)\psi
+g\left[(\overline{\psi}\psi)^{2}+(\overline{\psi}i\gamma_{5}
\tau^{a} \psi)^{2} \right],
\label{NJLmodel}
\end{equation}
where $g$ is a coupling constant, $\tau^{a}$ is the Pauli matrix
and $\mbox{\boldmath $m$}$ is a mass matrix.
This model takes into account the axial anomaly and
has $\mathrm{U_{V}}(1)\otimes \mathrm{SU_{V}}(2)\otimes \mathrm{SU_{A}}(2)$
chiral symmetry except for the mass term.

A remarkable features in the expected phase structure is the possible existence
of the first-order phase boundary between the hadronic phase
and the QGP phase at large baryon chemical potential $\mu$.
In particular, the phase transition becomes second order at
the end point of the first-order phase boundary, which is referred to the QCD CP.

In general, a system near the CP
shows large fluctuations of and correlations between various quantities
and thus a method beyond the
mean-field theory is desirable for describing the physical properties near the CP.
The FRG is expected to be a method to reveal the nature of the system more accurately than the mean-field theory
and has been found to be useful in the description of chiral phase
transition in QCD via effective chiral models
\cite{Jungnickel:1995fp,Braun:2003ii,Schaefer:2005,Schaefer:2006ds,
Stokic:2010, Nakano:2010, Aoki:2014}.
Moreover, there exist specific collective modes  which are coupled to the
fluctuations of the order parameter and become gapless and a long-life at
the CP. Such a mode is called the {\em soft mode} of the phase transition.
As for the QCD CP, the nature of the soft modes is nontrivial due to the current quark mass \cite{Fujii:2004jt,Son:2004iv}.
In the case of finite current quark mass, the universality class
of the CP belongs to that of $Z_{2}$ CP and the soft mode is
considered to be the particle-hole mode corresponding to the density (and energy) fluctuations. It is noteworthy that
the scalar-vector coupling\cite{Kunhiro:1991} caused
by the finite quark mass at nonvanishing $\mu$  leads to
a singular behavior of not only the chiral susceptibility but also
susceptibilities of  the hydrodynamical modes such as the density
fluctuation or the quark-number susceptibility at the CP,
as was shown in some model calculations
\cite{Fujii:2004jt,Son:2004iv}.
In Ref.~\cite{Yokota:2016tip},
FRG has been applied to calculate the spectral functions of the sigma meson and pion channels,
and thus the nature of the soft mode at the QCD CP was clarified.

In this lecture note,
 which is essentially a rearrangement of Ref.~\cite{Yokota:2016tip} with a focus on the technical part,
we show the way to calculate the meson spectral functions in the two-flavor quark-meson model with FRG
and its application to the analysis of the soft mode at the QCD CP.
Our results confirm the softening of the particle-hole mode
in the $\sigma$ channel near the QCD
CP, but not in the pion channel.
In addition, we find that the low-momentum dispersion
relation of sigma-mesonic mode penetrates into space-like region and the
mode merges into the bump of the particle-hole mode.

This note is organized as follows.
In Sec. \ref{SMethod}, we recapitulate the method
developed in \cite{Tripolt:2013jra,Tripolt:2014wra} and
describe details for numerical calculation.
The results are shown in
Sec. \ref{SNumericalResults}.
The phase diagram, the critical region and the precise location of the CP are presented in Sec. \ref{SSPhaseDiagram}.
In Sec. \ref{SSSpectral}
the results of the spectral functions are
shown, and  the soft mode at the QCD CP is discussed.
Sec. \ref{SSummary} is devoted to  summary and outlook.

\section{Method} \label{SMethod}

In this section, we summarize the method to calculate
the spectral functions in the FRG approach following Ref.\cite{Tripolt:2013jra,Tripolt:2014wra},
 and present a numerical
stability condition \cite{Yokota:2016tip} for solving the flow equation as an evolution equation.
The method is applied to the two-flavor quark-meson model.

\subsection{Procedure to derive spectral functions in meson channels}

\begin{figure}[!b]
\centering\includegraphics[width=0.9\columnwidth]{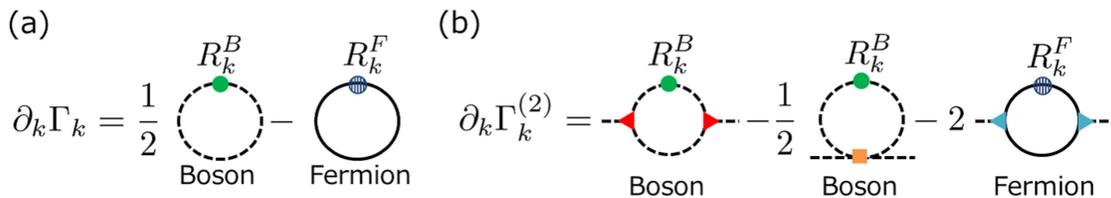}
\caption{Diagrammatic representations of (a) Eq. (\ref{Weq})
and (b) Eq. (\ref{G2flow}).}
\label{DiagRep}
\end{figure}

The FRG is based on the philosophy of the Wilsonian renormalization group
\cite{Wetterich:1992yh,Wegner:1972ih,Wilson:1974,Polchinski:1983gv}
and realizes the coarse graining by introducing a regulator function
$R_{k}$, which has a role to suppress modes with lower momentum  than the scale $k$ for the respective field. In this method, the effective
average action (EAA) $\Gamma_{k}$ is introduced such that it becomes bare action
$S_{\Lambda}$ at a large UV scale $k=\Lambda$ and becomes the effective
action at $k\rightarrow 0$ with an appropriate choice of
regulators. The flow equation for EAA, the Wetterich equation,
can be derived as a functional differential equation \cite{Wetterich:1992yh}:
\begin{equation}
 \partial_{k}\Gamma_{k}=\frac{1}{2}\mathrm{STr}\left[
\frac{\partial_{k}R_{k}}{\Gamma^{(2)}_{k}+R_{k}}
\right],
\label{Weq}
\end{equation}
where $\Gamma_{k}^{(n)}$ is the $n$-th functional derivative of
$\Gamma_{k}$ with respect to fields. This equation has a one-loop structure
and can be represented diagrammatically as shown in Fig. \ref{DiagRep} (a).
In principle, one can get the effective action $\Gamma_{k=0}$ by solving Eq. (\ref{Weq})
with the initial condition $\Gamma_{\Lambda}=S_{\Lambda}$.

The spectral function $\rho(\omega, p)$ for some field is derived from the imaginary part of the two-point retarded Green's function $G^{R}(\omega, p)$ for the field in momentum space:
\begin{equation}
\rho(\omega, p)=-\frac{1}{\pi}\mathrm{Im} G^{R}(\omega, p).
\label{rhoImG}
\end{equation}
Let us denote fields in a system as $\left\lbrace\varphi_{i} \right\rbrace$.
Suppose that the average of $\left\lbrace\varphi_{i} \right\rbrace$, denoted by $\left\lbrace\varphi_{i0} \right\rbrace$, is obtained.
Then, the inverse of the second derivative of the effective action at $\left\lbrace\varphi_{i}=\varphi_{i0} \right\rbrace$ leads to some two-point Green's function.
In the imaginary-time formalism, the inverse
of the second derivative of the effective action at
$\left\lbrace\varphi_{i}=\varphi_{i0}\right\rbrace$ is the Matsubara Green's function
and the analytic continuation from imaginary  to real frequency in the momentum space
gives the retarded Green's function if it retains the analyticity
of the Green's function in the upper half-plane of complex frequency \cite{BM1961}.
Therefore, one can get spectral functions by calculating second derivatives of EAA at $k=0$ using Eq. (\ref{Weq}).
The flow equation for the second derivative of EAA is derived from Eq. (\ref{Weq}) as
\begin{align}
\partial_{k}\Gamma^{(2)}_{k,ij}[\varphi_{0}]
=&
\left. \mathrm{STr}
\left[
\left[\Gamma^{(2)}_{k}+R_{k}\right]^{-1}_{ab}
\Gamma^{(3)}_{k,bci}
\left[\Gamma^{(2)}_{k}+R_{k}\right]^{-1}_{cd}
\Gamma^{(3)}_{k,dej}
\left[\Gamma^{(2)}_{k}+R_{k}\right]^{-1}_{ef}
\partial_{k}R_{k}^{fa}
\right]
\right|_{\varphi=\varphi_{0}}\notag \\
&-
\left. \frac{1}{2}
\mathrm{STr}
\left[
\left[\Gamma^{(2)}_{k}+R_{k}\right]^{-1}_{ab}
\Gamma^{(4)}_{k,bcij}
\left[\Gamma^{(2)}_{k}+R_{k}\right]^{-1}_{cd}
\partial_{k}R_{k}^{da}
\right]
\right|_{\varphi=\varphi_{0}},
\label{G2flow}
\end{align}
where $\Gamma^{(n)}_{k,i_{1}\cdots i_{n}}$ represents the $n$th derivative of $\Gamma_{k}$
with respect to $\varphi_{i_{1}}, \cdots ,\varphi_{i_{n}}$.
The diagrammatic expression of this equation is shown in Fig. \ref{DiagRep} (b).
The RHS of Eq. (\ref{G2flow}) contains $\Gamma_{k}^{(2)}$, $\Gamma_{k}^{(3)}$ and $\Gamma_{k}^{(4)}$.
In general the flow equation consists of an infinite hierarchy of differential equations
such that the flow equation for $\Gamma_{k}^{(n)}$ contains
$\Gamma_{k}^{(n+1)}$ and $\Gamma_{k}^{(n+2)}$.
Some simplification of this hierarchy is needed to solve Eq. (\ref{G2flow}).
One of the simplifications is evaluating the RHS of Eq. (\ref{G2flow}) with a truncated EAA and
integrating the equation to $k=0$. Adopting this simplification, we
present the calculation
of the meson spectral functions in the quark-meson model below.

We employ the two-flavor quark-meson model as the low-energy effective model of QCD. This is a chiral effective model consisting of the quark field and auxiliary fields $\sigma$ and $\vec{\pi}$ corresponding to $\overline{q}q$ and $\overline{q}i\vec{\tau}\gamma_{5}q$, respectively.
We analyse in finite temperature and chemical potential. The bare action for this model in imaginary-time formalism is as follows:
\begin{equation}
S_{\Lambda}\left[\pbar,\psi, \phi \right]
=
\int_{0}^{\frac{1}{T}}d\tau
\int d^{3}\vec{x}
\left\lbrace
\pbar
\left(
\Slash{\partial}
+g_{s}(\sigma+i\vec{\tau}\cdot\vec{\pi}\gamma_{5})
-\mu \gamma_0
\right)
\psi
+\frac{1}{2}(\partial_{\mu}\phi)^2
+V(\phi^2)-c\sigma
\right\rbrace,
\label{eq:meson-quark-model}
\end{equation}
where $\phi=(\sigma,\vec{\pi})$. The quark field $\psi$ has the indices
of four-component spinor, color $N_{c}=3$ and flavor $N_{f}=2$.
The last term $c\sigma$ represents the effect of the current quark mass, which
explicitly breaks chiral symmetry.
$V(\phi^{2})$ is the potential term of the mesons. We remark that
the essential part of the meson-quark model (\ref{eq:meson-quark-model}) may be obtained
 by a Hubbard-Stratonovich transformation of the NJL model (\ref{NJLmodel}).

Now we take the local
potential approximation (LPA) for the meson flow part as our
truncation scheme. This truncation corresponds to considering only
the lowest-order of derivative expansion for the meson flow part.
Our truncated EAA is as follows
\cite{Schaefer:2006ds}:
\begin{align}
\Gamma_{k}\left[\pbar,\psi,\phi \right]
=&
\int_{0}^{\frac{1}{T}}d\tau
\int d^{3}x
\left\lbrace
\pbar
\left(\Slash{\partial}
+g_{s}(\sigma+i\vec{\tau}\cdot\vec{\pi}\gamma_{5})
-\mu \gamma_{0} \right)
\psi
+\frac{1}{2}(\partial_{\mu}\phi)^2
+U_{k}(\phi^2)-c\sigma
\right
\rbrace,
\label{QMtranc}
\end{align}
where $U_{k}(\phi^{2})$ satisfies $U_{\Lambda}(\phi^{2})=V(\phi^{2})$.
In this truncation, we also
neglect the flow of $g_s$ and the wave function renormalization. Therefore,
only the meson effective potential $U_{k}$ has a $k$-dependence.
The nonperturbative effects are to be incorporated through Eq. (\ref{G2flow})
with the truncated EAA used as the initial condition.

The procedure to calculate two-point functions is as follows.
We first calculate the effective potential $U_{k}(\phi^{2})$ using
Eq. (\ref{Weq}). Then the chiral condensate $\sigma_{0}$, i.e., the average of $\sigma$, is obtained
as $\sigma$ satisfying the quantum equation of motion
(EOM) $\delta \Gamma_{k=0}/\delta \sigma = 0$. In our case,
this condition corresponds to obtaining $\sigma$ that minimizes
$U_{k}(\sigma^{2})-c\sigma$,
under the assumptions that the condensate is homogeneous
and $\langle \vec{\pi} \rangle = \vec{0}$.

Next, we derive the flow equations for two-point Green's functions from Eq. (\ref{G2flow}).
We define $\mathscr{G}_{k,\sigma}(P)$ and $\mathscr{G}_{k,\sigma}(P)$ as
\begin{align}
\left.
\frac{\delta^{2} \Gamma_{k}}{\delta \sigma(\pfour) \delta \sigma(\qfour)}
\right|
_{\pbar=0,\psi=0,\vec{\pi}=\vec{0},\sigma=\sigma_{0}}
&=(2\pi)^{4}\delta^{(4)}(\pfour + \qfour)\mathscr{G}^{-1}_{k,\sigma}(P), \\
\left.
\frac{\delta^{2} \Gamma_{k}}{\delta \pi_{a}(\pfour) \delta \pi_{a}(\qfour)}
\right|
_{\pbar=0,\psi=0,\vec{\pi}=\vec{0},\sigma=\sigma_{0}}
&=(2\pi)^{4}\delta^{(4)}(\pfour + \qfour)\mathscr{G}^{-1}_{k,\pi}(P),
\end{align}
where $\sigma(P)$ and $\pi_{a}(P)$ are the momentum space representations of the sigma and pion fields, respectively,
and $\pfour=(i\omega_{n},\vec{p})$ with
$\omega_{n}$ being the bosonic Matsubara frequency.
These quantities become Matsubara Green's functions at $k=0$.
By inserting the average values $\sigma=\sigma_{0}$, $\vec{\pi}=\vec{0}$, and $\pbar=\psi=0$ and choosing $\varphi_{i}=\sigma(P)$ and $\varphi_{j}=\sigma(Q)$ ($\varphi_{i}=\pi_{a}(P)$ and $\varphi_{j}=\pi_{a}(Q)$) in Eq. (\ref{G2flow}), the flow equation for $\mathscr{G}_{k,\sigma}(P)$ ($\mathscr{G}_{k,\pi}(P)$) can be obtained.
In our approximation, the $\Gamma_{k}^{(2)}$, $\Gamma_{k}^{(3)}$ and $\Gamma_{k}^{(4)}$
in the RHS of Eq. (\ref{G2flow}) are evaluated using Eq. (\ref{QMtranc}).

We define $G^{R}_{k,\sigma}(P)$ ($G^{R}_{k,\pi}(P)$)
as the function which is obtained by analytic continuation for $\mathscr{G}_{k,\sigma}(P)$ ($\mathscr{G}_{k,\pi}(P)$) from imaginary Matsubara frequencies to real frequencies retaining
the analyticity in the upper half-plane of complex frequency.
The solutions of $G^{R}_{k,\sigma}(P)$ and $G^{R}_{k,\pi}(P)$ at $k=0$ give the retarded Green's functions in the
sigma and pion channel, respectively. Therefore, if analytic continuation of the flow equations for
$\mathscr{G}_{k,\sigma}(P)$ ($\mathscr{G}_{k,\pi}(P)$) is performed and the flow equation
for $G^{R}_{k,\sigma}(P)$ ($G^{R}_{k,\pi}(P)$) is obtained,
one can calculate the retarded Green's function using the flow equation.
In our case, such an analytic continuation is successfully carried as follows:
As mentioned above,
the analyticity of the Green's function in the upper half-plane of $\omega$ must be retained.
In the present case, the flow equation itself should be analytic
in the upper half-plane after the analytic continuation.
One can retain the analyticity in the upper half-plane
easily by taking into account the following points.
\begin{enumerate}
\item By choosing $\omega_n$-independent regulators,
one can avoid possible extra poles in the $\omega$ plane in the flow equation otherwise arising from $\omega_n$ dependence of the regulators.
\item The next point is about the analytic continuation of thermal
distribution functions $n_{B,F}(E+i\omega_{n})$ obtained for a discrete (multiple of $2\pi T$)
frequency $\omega_n$, where
the subscript $B$, $F$ stands for a boson or fermion, respectively, and $E$ is $\omega_{n}$ independent.
Such factors appear in the flow equation after the
Matsubara summation.
Because of the periodicity of the exponential function,
$n_{B,F}(E+i\omega_{n})$ is equal to $n_{B,F}(E)$.
However if $n_{B,F}(E+\omega)$
is substituted for $n_{B,F}(E+i\omega_{n})$,
such a factor breaks the analyticity of the flow equation
in the upper half-plane.
Therefore $n_{B,F}(E+i\omega_{n})$ should be
replaced by $n_{B,F}(E)$ before the analytic continuation.
\end{enumerate}
By taking into account these points,
the substitution $\omega+i\epsilon$ for $i\omega_{n}$
with $\epsilon$ being a positive infinitesimal gives
the flow equations for $G^{R}_{k,\sigma}(P)$ and $G^{R}_{k,\pi}(P)$.

Finally, the spectral functions in the meson channels are given in terms of the thus-obtained retarded Green's functions $G^{R}_{k\rightarrow 0,\sigma}(P)$ and $G^{R}_{k\rightarrow 0,\pi}(P)$ using Eq. (\ref{rhoImG}).

\subsection{Flow equations}
In the present work, we adopt the 3D Litim's optimized
regulators for bosons and fermions \cite{Litim:2001up}
as $\omega_{n}$-independent regulators:
\begin{eqnarray}
\RB(\qfour)&=&(k^2-\vec{q}^2)\theta(k^2-\vec{q}^2),\label{eq:reg_boson} \\
\RF(\qfour)&=&i\Slash{\vec{q}}\left(\sqrt{\frac{k^2}{\vec{q}^2}}-1\right)\theta(k^2-\vec{q}^2).\label{eq:reg_fermion}
\end{eqnarray}
Then the insertion  of Eq. (\ref{QMtranc})
into Eq. (\ref{Weq}) leads to the following flow equation for $U_{k}$:
\begin{align}
\partial_{t}U_k
=\frac{k^5}{12\pi^2}
&\left[
-2N_f N_c
\left[
\frac{1}{E_{\psi}}
\tanh\frac{E_{\psi}+\mu}{2T}
+
\frac{1}{E_{\psi}}
\tanh\frac{E_{\psi}-\mu}{2T}
\right]
+
\frac{1}{E_{\sigma}}
\coth\frac{E_{\sigma}}{2T}
+
\frac{3}{E_\pi}
\coth\frac{E_\pi}{2T}
\right],
\label{Ukflow}
\end{align}
where $t=\ln(k/\Lambda)$, $E_{a}=\sqrt{k^2+m^2_{a}}\ (a=\psi ,\sigma,\pi)$, and
\begin{align}
m^2_{\psi}=g^2_s \sigma^2,\  m^2_{\sigma} =\partial^2_{\sigma}U_k,\  m^2_{\pi} =\partial_\sigma U_k /\sigma .
\label{Masses}
\end{align}

According to the procedure presented in the previous subsection,
the flow equations for $\Gamma^{(2)}_{k,\sigma}(\pfour)$ and
$\Gamma^{(2)}_{k,\pi}(\pfour)$ become:
\begin{align}
\Dk\Gs2(\pfour)
=&
\Jss(\pfour)(\Gamma^{(0,3)}_{k,\sigma\sigma\sigma})^{2}
-\frac{1}{2}\Is \Gamma^{(0,4)}_{k,\sigma\sigma\sigma\sigma}
+3\Jpp(\pfour)(\Gamma^{(0,3)}_{k,\sigma\pi\pi})^{2}
-\frac{3}{2}\Ip \Gamma^{(0,4)}_{k,\sigma\sigma\pi\pi}
-2N_{c}N_{f}\Jsqq(\pfour),
\label{G2Sflow} \\
\Dk\Gp2(\pfour)
=&
\Jsp(\pfour)(\Gamma^{(0,3)}_{k,\sigma\pi\pi})^{2}+\Jps(\pfour)(\Gamma^{(0,3)}_{k,\sigma\pi\pi})^{2}
-\frac{1}{2}\Is\Gamma^{(0,4)}_{k,\sigma\sigma\pi\pi}
-\frac{5}{2}\Ip\Gamma^{(0,4)}_{k,\pi\pi\tilde{\pi}\tilde{\pi}} -2N_{c}N_{f}\Jpqq(\pfour),
\label{G2Pflow}
\end{align}
respectively, where $\pi,\tilde{\pi} \in \lbrace \pi_{1},\pi_{2},\pi_{3}\rbrace$ and $\pi\neq \tilde{\pi}$.
The loop-functions
$J_{k,\alpha\beta}(P)$,\, $I^{(2)}_{k,\alpha}$, and $J_{k,\pbar \psi}^{(\alpha)}(P)$\, $(\alpha,\beta=\sigma,\pi)$ are defined as
\begin{align}
J_{k,\alpha\beta}(\pfour)&=T\sum_{q_{n}}\int \frac{d^{3}\vec{q}}{(2\pi)^{3}} \partial_{k} R_{k}^{B}(q)
G^{B}_{k,\alpha}(P)^{2} G^{B}_{k,\beta}(Q-P),
\label{Jab}
\\
I^{(2)}_{k,\alpha}&=T\sum_{q_{n}}\int \frac{d^{3}\vec{q}}{(2\pi)^{3}} \partial_{k} R_{k}^{B}(q)
G^{B}_{k,\alpha}(Q)^{2},
\label{Ia}
\\
J_{k,\pbar \psi}^{(\alpha)}(\pfour)&=T\sum_{q_{n}}\int \frac{d^{3}\vec{q}}{(2\pi)^{3}} \mathrm{tr}
\left[
\Gamma^{(2,1)}_{\pbar\psi\alpha}
G^{F}_{k,\pbar \psi}(Q)
\partial_{k}R_{k}^{F}(Q)
G^{F}_{k,\pbar \psi}(Q)
\Gamma^{(2,1)}_{\pbar\psi\alpha}
G^{F}_{k,\pbar \psi}(Q-P)
\right],
\label{Jqq}
\end{align}
where $\qfour=(iq_{n},\vec{q})$ and
\begin{align}
G^{B}_{k,\alpha}(\qfour)&=\left[\qfour^{2}+\left. m_{\alpha}^{2}\right|_{\sigma=\sigma_{0}}+R_{k}^{B}(\qfour)
\right]^{-1},
\\
G^{F}_{k,\pbar \psi}(\qfour)&=
\left[\Slash{\qfour}-\mu \gamma_{0}+\left. m_{\psi}\right|_{\sigma=\sigma_{0}}+R^{F}_{k}(\qfour)\right]^{-1}.
\end{align}
The three- and four-point vertices $\Gamma^{(2,1)}_{\pbar\psi\phi_{i}}$,\,
$\Gamma^{(0,3)}_{k,\phi_{i}\phi_{j}\phi_{l}}$,
and $\Gamma^{(0,4)}_{k,\phi_{i}\phi_{j}\phi_{l}\phi_{m}}$
are defined as
\begin{align}
\frac{\delta}{\delta \phi_{i} (P_{1})}
\frac{\overset{\rightarrow}{\delta}}{\delta \pbar (P_{2})}
\Gamma_{k}
\frac{\overset{\leftarrow}{\delta}}{\delta \psi (P_{3})}
&=(2\pi)^{4}\delta^{(4)}(P_{1}+P_{2}+P_{3})
\Gamma^{(2,1)}_{\pbar\psi\phi_{i}},
\\
\frac{\delta^{3} \Gamma_{k}}{
\delta \phi_{i} (P_{1})
\delta \phi_{j} (P_{2})
\delta \phi_{l} (P_{3})}
&=(2\pi)^{4}\delta^{(4)}(P_{1}+P_{2}+P_{3})
\Gamma^{(0,3)}_{k,\phi_{i}\phi_{j}\phi_{l}}, \\
\frac{\delta^{4} \Gamma_{k}}{
\delta \phi_{i} (P_{1})
\delta \phi_{j} (P_{2})
\delta \phi_{l} (P_{3})
\delta \phi_{m} (P_{4})}
&=(2\pi)^{4}\delta^{(4)}(P_{1}+P_{2}+P_{3}+P_{4})
\Gamma^{(0,4)}_{k,\phi_{i}\phi_{j}\phi_{l}\phi_{m}},
\end{align}
some of which are expressed in terms of $U_{k}$:
\begin{align}
\Gamma^{(2,1)}_{\pbar\psi\phi_{i}}
&=
\begin{cases}
g_{s} &(\text{for }i=0)\\
g_{s}i\gamma^{5}\tau^{i} &(\text{for }i=1,2,3)
\end{cases},
\\
\Gamma^{(0,3)}_{k,\phi_{i}\phi_{j}\phi_{l}}
&=4U_{k}^{(2)}
(\delta_{ij}\phi_{m}+\delta_{im}\phi_{j}+\delta_{jm}\phi_{i})
+8U_{k}^{(3)}\phi_{i}\phi_{j}\phi_{m},
\\
\Gamma^{(0,4)}_{k,\phi_{i}\phi_{j}\phi_{l}\phi_{m}}
&=4U_{k}^{(3)}
(\delta_{ij}\delta_{mn}+\delta_{in}\delta_{jm}+\delta_{jn}\delta_{im}) \notag \\
&+8U_{k}^{(3)}(
\delta_{ij}\phi_{l}\phi_{m}
+\delta_{jl}\phi_{i}\phi_{m}
+\delta_{lm}\phi_{i}\phi_{j}
+\delta_{jm}\phi_{i}\phi_{l}
+\delta_{im}\phi_{j}\phi_{l}
+\delta_{il}\phi_{j}\phi_{m}
) \notag \\
&+16U_{k}^{(4)}\phi_{i}\phi_{j}\phi_{l}\phi_{m}.
\end{align}
Analytic continuation in Eq. (\ref{G2Sflow}) and Eq. (\ref{G2Pflow})
is carried out after the Matsubara summation in Eqs. (\ref{Jab})--(\ref{Jqq}).
The explicit forms of Eq. (\ref{Jab})
- (\ref{Jqq}) after Matsubara summation is shown in \cite{Yokota:2016tip}.

To solve these flow equations,
we employ the following initial conditions at the UV scale $k=\Lambda$:
\begin{align}
&U_{\Lambda}(\phi^2)=V(\phi^2)=\frac{1}{2}m_{\Lambda}^2\phi^2+\frac{1}{4}\lambda_{\Lambda}(\phi^2)^2, \label{initial}
\\
&\Gamma_{\Lambda ,\sigma}^{(2),R}(\omega,\vec{p})
=-\omega^{2}+\vec{p}^{2}+\partial^2_{\sigma}U_{\Lambda}(\sigma_{0}^{2}),  \\
&\Gamma_{\Lambda ,\pi}^{(2),R}(\omega,\vec{p})
=-\omega^{2}+\vec{p}^{2}+\partial_\sigma U_{\Lambda}(\sigma_{0}^{2}) /\sigma_{0}.
\end{align}

\subsection{Numerical stability conditions}
\label{AppStability}
We employ the grid method to solve Eq. (\ref{Ukflow}) numerically.
This method reveals the global structure of $U_{k}(\sigma^{2})$ on
discretized $\sigma$. We employ the fourth-order Runge-Kutta method to solve Eq. (\ref{Ukflow}).

In general when one solves a partial differential equation
numerically, the discretization of derivatives may cause numerical
errors. Thus, one needs to impose numerical stability conditions to
avoid the enhancement of the error due to accumulation.
The derivation of such conditions is concretely demonstrated
in the case of linear partial differential equations and briefly mentioned in the case of nonlinear partial
differential equations in \cite{NumericalRecipes}.

A numerical stability conditions for numerical calculation
was given for the following partial differential equation
in \cite{Yokota:2016tip}:
\begin{equation}
\frac{\partial u(t,\sigma)}{\partial t}=f\left(t,\sigma,u(t,\sigma), \frac{\partial u(t,\sigma)}{\partial \sigma},\frac{\partial^{2}u(t,\sigma)}{\partial \sigma^{2}}\right).
\label{flow}
\end{equation}
where $f$ is an arbitrary real function.
This equation is a generalized equation of Eq. (\ref{Ukflow}).
The equation to describe the evolution of numerical deviation
for $u(t,\sigma)$ can be derived from the descritized form of Eq. (\ref{flow}) for $t$ and $\sigma$, from which one can derive the conditions for suppressing the amplification of the numerical deviation, i.e., the numerical stability conditions.
If we choose forward difference for $t$-derivative
and central three-point difference for $\sigma$-derivative as
the descritization the stability conditions for Eq. (\ref{flow}) are as follows \cite{Yokota:2016tip}:
\begin{numcases}
{}
|\Delta t| \leq \frac{2|G|}{F^{2}}, & \notag \\
|\Delta t| \leq \frac{\Delta \sigma^{2}}{2|G|},& \label{finalcond}
\end{numcases}
where
\begin{equation}
F\equiv\frac{\partial f}{\partial u'},\
G\equiv\frac{\partial f}{\partial u''}
\  \left(u'=\frac{\partial u}{\partial \sigma}, u''= \frac{\partial^{2} u}{\partial \sigma^{2}} \right). \notag
\end{equation}
In the case of Eq. (\ref{Ukflow}),
$t$ and $u(t,\sigma)$ are identified with $\ln(k/\Lambda)$ and $U_{k}(\sigma^{2})$, respectively,
and $F$ and $G$ are derived to be:
\begin{numcases}
{}
F=-\frac{k^{5}}{8\pi^{2}\sigma E_{\sigma}^{3}}
\left(\coth\frac{E_{\pi}}{2T}+\frac{E_{\pi}}{2T}\frac{1}{\sinh^{2}\frac{E_{\pi}}{2T}}\right), & \label{FQM}\\
G=-\frac{k^{5}}{24\pi^{2}E_{\sigma}^{3}}
\frac{\coth\frac{E_{\sigma}}{2T}+\frac{E_{\sigma}}{2T}\frac{1}{\sinh^{2}\frac{E_{\sigma}}{2T}}}
{\left(\coth\frac{E_{\pi}}{2T}+\frac{E_{\pi}}{2T}\frac{1}{\sinh^{2}\frac{E_{\pi}}{2T}}\right)^{2}}. & \label{GQM}
\end{numcases}
$G$ is negative definite, and $\Delta t$ is also negative
because the direction of flow is from $k=\Lambda$ to $k=0$.

We fix the intervals of discretization of $\sigma$ and $t$
in Eq. (\ref{Ukflow})
according to these conditions.
Because the above condition of Eq. (\ref{finalcond}) is
too strict when $\sigma$ is close to zero, we neglect
the condition around $\sigma=0$ in practice.

\subsection{Other numerical details}
As stated before, the flow equation (\ref{Ukflow}) should be
integrated down to $k=0$ from $k=\Lambda$ to get the effective action $\Gamma_{k=0}$ in principle.
However, due to the conditions for stable calculation mentioned above,
solving the flow equation to small $k$ is quite time-consuming
for some regions of the $(T,\mu)$ plane, such as the
low-temperature region of the hadronic phase.
In such a region, the curvature of $U_k$, i.e., $m^2_\sigma$, can take a negative value, which leads to small $E_\sigma$ for some $\sigma$. This gives large $F$ and $|G|$ (Eqs.(\ref{FQM}) and (\ref{GQM})) as $k$ decreases and the condition (\ref{finalcond}) becomes difficult to satisfy at small $k$. Thus, some infrared scale $k=k_{\mathrm{IR}}$ is introduced in practice, at which the numerical procedure is stopped. Of course, $k_{\mathrm{IR}}$ should be as small as possible so that sufficiently low-momentum fluctuations are taken into account
to describe the system around the CP where vanishingly low-momentum excitations exist. Thus we choose a much lower value of $k_{\mathrm{IR}}$ than the $40\,\MeV$ adopted in Ref. \cite{Tripolt:2014wra}, and set $k_{\mathrm{IR}}=1\,\MeV$ as being small enough to incorporate the low-momentum fluctuations.
Therefore our calculation will be reliable
in the vicinity of the CP,
except for the small surrounding region where
excitation modes with momentum scales lower than $1\,\MeV$
are strongly developed.
Although $\epsilon$, which appears after the analytic
continuation, is defined as a positive infinitesimal,
we set it to $1\,\MeV$ in the present calculation, which should be small enough for present purposes.

3D momentum integrals remain
after the Matsubara summation in Eq. (\ref{G2Sflow}) and Eq. (\ref{G2Pflow}).
These integrals can be fully calculated analytically
for zero external momentum.
Even for a finite external momentum,
they can be nicely reduced to 1D integrals, which are evaluated numerically.
The numerical integrations involve a tricky point, and one has to take care of
the poles of each term in the integrands.
As an example, we show the explicit form of $J_{k,\alpha\beta}(P)$ after Matsubara summation:
\begin{align}
\Jab(\pfour)
=&\int_{D_{1}} \frac{d^3 q}{\left (2\pi\right )^{3}}
\frac{k}{2}
\left[
(1+n_B(\Ea))\frac{\Ea^2+\Eb^2-(2\Ea+\ip0)^2}{\Ea^3(\Eb^2-(\Ea+\ip0)^2)^2}
+n_B(\Ea)\frac{\Ea^2+\Eb^2-(2\Ea-\ip0)^2}{\Ea^3(\Eb^2-(\Ea-\ip0)^2)^2}  \right. \notag \\
&+\frac{2(1+n_B(\Eb))}{\Eb(\Ea^2-(\Eb-\ip0)^2)^2}
+\frac{2n_B(\Eb)}{\Eb(\Ea^2-(\Eb+\ip0)^2)^2}
\left. -\frac{n_{B}^{\prime}(\Ea)}{\Ea^2(\Eb^2-(\Ea-\ip0)^2)}
-\frac{n_{B}^{\prime}(\Ea)}{\Ea^2(\Eb^2-(\Ea+\ip0)^2)} \right] \notag \\
&+\int_{D_{2}} \frac{d^3 q}{\left (2\pi\right )^{3}}
\frac{k}{2}\left[(1+n_B(\Ea))\frac{\Ea^2+\Ebt^2-(2\Ea+\ip0)^2}{\Ea^3(\Ebt^2-(\Ea+\ip0)^2)^2}
+n_B(\Ea)\frac{\Ea^2+\Ebt^2-(2\Ea-\ip0)^2}{\Ea^3(\Ebt^2-(\Ea-\ip0)^2)^2} \right. \notag \\
&+\frac{2(1+n_B(\Ebt))}{\Ebt(\Ea^2-(\Ebt-\ip0)^2)^2}
+\frac{2n_B(\Ebt)}{\Ebt(\Ea^2-(\Ebt+\ip0)^2)^2}
\left. -\frac{n_{B}^{\prime}(\Ea)}{\Ea^2(\Ebt^2-(\Ea-\ip0)^2)}
-\frac{n_{B}^{\prime}(\Ea)}{\Ea^2(\Ebt^2-(\Ea+\ip0)^2)} \right],
\label{Jabform}
\end{align}
where $E_{\alpha}=\sqrt{k^{2}+m_{\alpha}^{2}}$, $\tilde{E}_{\alpha}=\sqrt{\vec{q}^2+m_{\alpha}^{2}}$, $n_{B,F}^{\prime}(E)=dn_{B,F}(E)/dE$,
\begin{equation}
D_1=\left\lbrace
\left. \vec{q} \in \mathbb{R}^{3} \right|
|\vec{q}-\vec{p}|<k\ \text{and}\ |\vec{q}|<k
\right\rbrace,\ \
D_2=\left\lbrace
\left. \vec{q} \in \mathbb{R}^{3} \right|
|\vec{q}-\vec{p}|<k\ \text{and}\ |\vec{q}|>k
\right\rbrace,\notag
\end{equation}
and $ip_{0}=\omega + i\epsilon$.
The first  and the third terms of the integrand of the second integral in Eq. (\ref{Jabform})
have the same pole $\tilde{E}_{\alpha}=E_{\alpha}+ip_{0}$.
If such terms are integrated separately, a large cancellation
can occur, which then leads to big numerical errors.
Therefore,
we first combine such  terms analytically in the integrand
before numerical integrations.

\subsection{Parameter setting}

The truncated EAA Eq. (\ref{QMtranc}) and the initial condition
Eq. (\ref{initial}) have some parameters which are fixed so as to
reproduce the observables in vacuum:
We use the same values of the parameters as those in
\cite{Tripolt:2014wra} and list them in Table \ref{Parameters}.

The chiral condensate $\sigma_{0}$ is determined as $\sigma$ which
minimize $U_{k}(\sigma^{2})-c\sigma$ and the constituent quark mass
$M_{\psi}$ and the sigma and pion screening masses $M_{\sigma}$ and
 $M_{\pi}$ are calculated using Eq. (\ref{Masses}):
\begin{equation}
M_{a}=\left( \left. m^2_{a} \right|_{\sigma=\sigma_{0}, k=k_{\mathrm{IR}}}\right)^{\frac{1}{2}},
\quad \ \ \ (a=\psi, \sigma, \pi).
\end{equation}
Our parameters reproduce
$\sigma_{0}=93\mathrm{MeV}$, $M_{q}=286\mathrm{MeV}$, $M_{\pi}=137\mathrm{MeV}$ and
$M_{\sigma}=496\mathrm{MeV}$ in vacuum.

\begin{table}[!tb]
\begin{center}
\begin{tabular}{ccccc}
\toprule
$\Lambda$ & $m_{\Lambda}/\Lambda$ &
$\lambda_{\Lambda}$ & $c/\Lambda^{3}$ &
$g_{s}$ \\
\hline
1000$\mathrm{MeV}$ & 0.794 & 2.00 & 0.00175 & 3.2
\\
\bottomrule
\end{tabular}
\end{center}
\caption{Used values of $\Lambda$ and the parameters in the initial condition
$\Gamma_{k=\Lambda}$ used in the calculation.}
\label{Parameters}
\end{table}

\section{Results} \label{SNumericalResults}
\subsection{Phase diagram}
\label{SSPhaseDiagram}

\begin{figure}[!tb]
\centering\includegraphics[width=0.95\columnwidth]{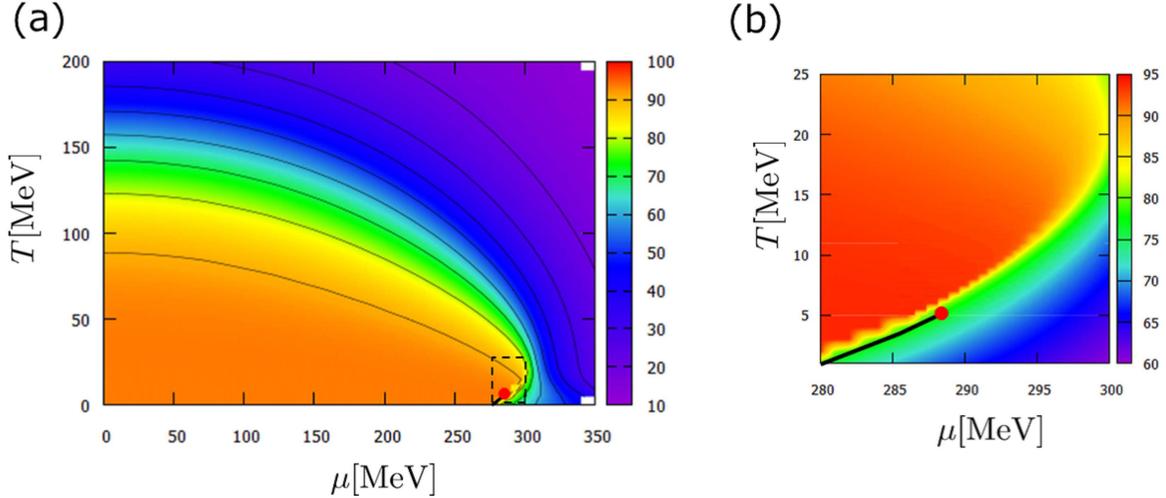}
\caption{(a) Contour map of the chiral condensate $\sigma_{0}$. (b) Enlargement of the area surrounded
by dotted lines in (a). The filled circle (red) is the CP
and the solid (black) line is the first-order phase boundary.
(Taken from  \cite{Yokota:2016tip})}
\label{phase}
\end{figure}

We show the phase diagram on temperature ($T$) and chemical potential ($\mu$) plane in Fig. \ref{phase}, where a contour map of the chiral condensate is also given.
One sees that chiral restoration occurs as the temperature is raised,
and the phase transition is not a genuine one but a crossover, except for
the low-temperature and large chemical potential region, where the phase
transition is of first order. This feature is qualitatively in accordance with
the results given in the literature, although the location of the CP here is in a somewhat smaller temperature region than that given in Ref. \cite{Tripolt:2014wra}.
The detailed procedure for locating the CP is described below.

\begin{figure}[!tb]
\centering\includegraphics[width=0.85\columnwidth]{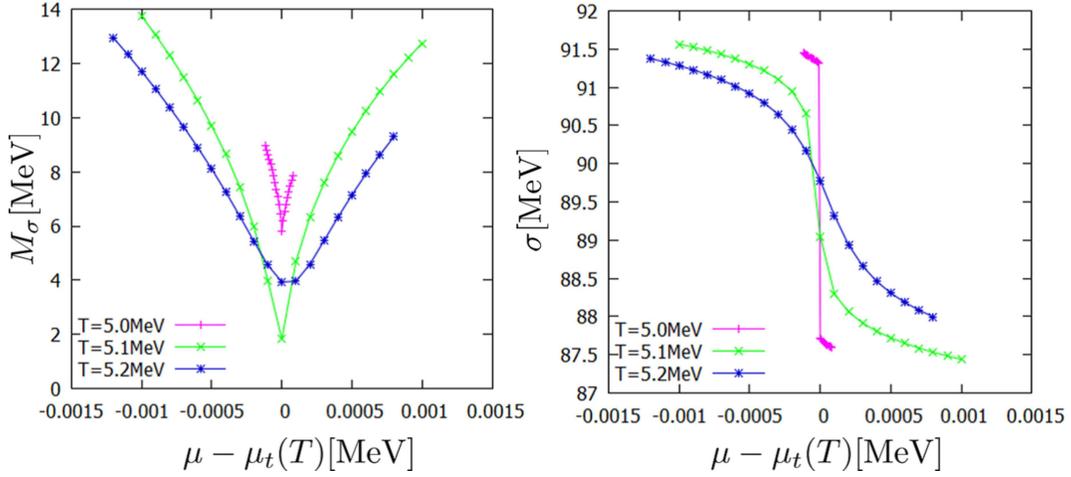}
\caption{The sigma curvature masses $M_{\sigma}$ and the chiral condensates $\sigma_{0}$ at $T=5.0\,\MeV$, $5.1\,\MeV$, and $5.2\,\MeV$. (Taken from  \cite{Yokota:2016tip})}
\label{Ms}
\end{figure}

At the QCD CP, the chiral susceptibility diverges.
Therefore, we locate the CP by searching for the point where the
sigma screening mass $M_{\sigma}$, the square of which is the inverse of the chiral susceptibility,
becomes the smallest: We seek the minimum position of $M_{\sigma}$ using the data points where
$M_{\sigma}$ is greater than $1\,\MeV$,
because our choice of $k_{\mathrm{IR}}=1\,\MeV$ enables us
to take into account fluctuations whose momentum scales are
greater than $k_{\mathrm{IR}}$ so as to make the result of $M_{\sigma}$ reliable when $M_{\sigma}$ is larger than $1\,\MeV$.
We also identify the first-order phase transition by
a discontinuity of the chiral condensate. The results at
$T=5.0\,\mathrm{MeV}$, $5.1\,\mathrm{MeV}$, and $5.2\,\mathrm{MeV}$
are shown in Fig. \ref{Ms} as functions of $\mu-\mu_{t} (T)$,
where $\mu_{t} (T)$ is the transition chemical potential
for each temperature determined by the minimum point of the sigma curvature mass and is found to be
$\mu_{t}(5.0\,\MeV)=286.517~02\,\MeV$,
$\mu_{t}(5.1\,\MeV)=286.686~00\,\MeV$, and
$\mu_{t}(5.2\,\MeV)=286.853~20\,\MeV$.

We find that the sigma screening mass becomes smallest between
$T=5.0\,\MeV$ and $T=5.2\,\MeV$ and between $\mu=\mu_{t}(5.0\,\MeV)$ and
$\mu=\mu_{t}(5.2\,\MeV)$. Therefore, the critical temperature $T_{c}$ and
the critical chemical potential $\mu_{c}$ are estimated as
$T_{c}=5.1\pm 0.1\,\MeV$ and $\mu_{c}=286.6 \pm 0.2\,\MeV$.
The position of the CP is quite different from  the $(T,\mu)=(10\,\MeV, 292.97\,\MeV)$ given in Ref. \cite{Tripolt:2014wra}.
Such a difference may be attributed to the different choice of
$k_{\mathrm{IR}}$.
In the following discussion,
we regard $T_{c}$ and $\mu_c$ as $5.1\,\MeV$ and
 $286.686\,\MeV$, respectively.
As seen in the behavior of the chiral condensate shown in the
right panel of Fig. \ref{Ms}, the
phase transition along the chemical potential is of first order when $T=5.0\,\MeV$ and a crossover when $T=5.2\,\MeV$.

\subsection{Examples of the results of the spectral function in the $\sigma$ channel}
\label{SSExample}

We first show examples of the results of
the spectral function in the $\sigma$ channel $\rho_{\sigma}(\omega, p)$.
\begin{figure}[!t]
\centering\includegraphics[width=1.0\columnwidth]{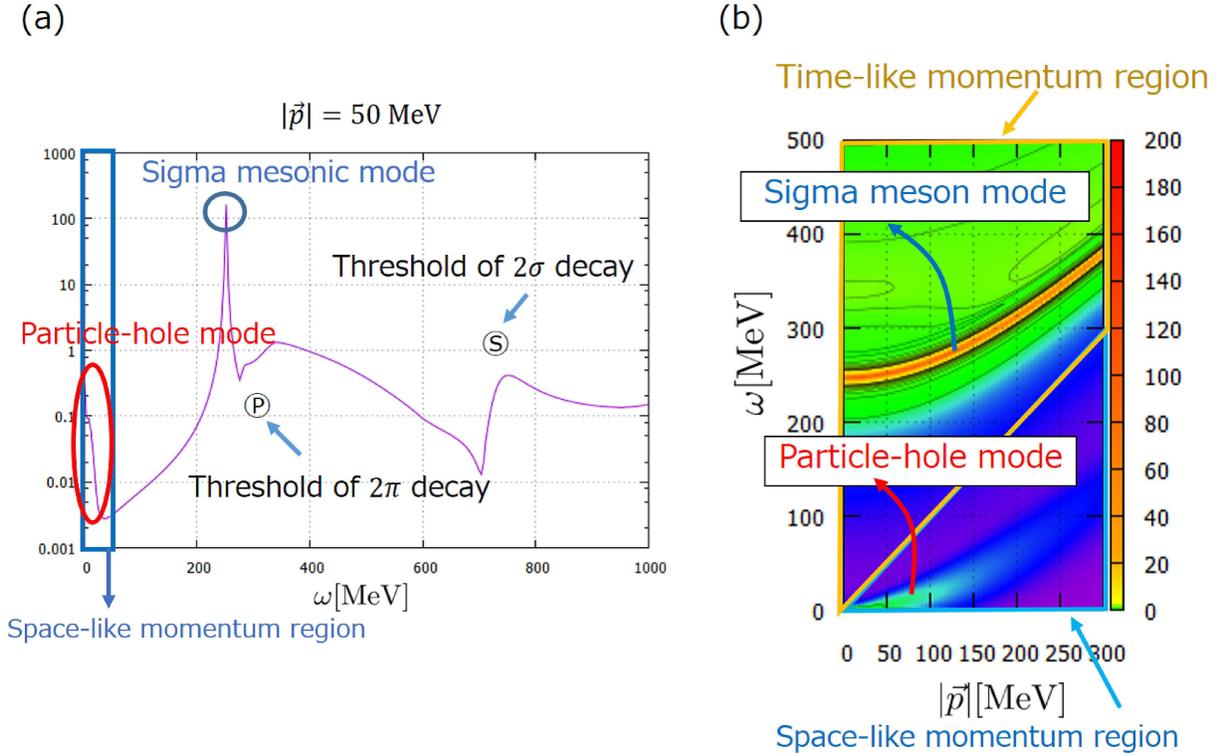}
\caption{An example of results of $\rho_{\sigma}(\omega, p)$:
(a) the result when $|\vec{p}|=50\MeV$ at $(T, \mu)=(5.1\MeV, 286.3 \MeV)$ and (b) the contour map of $\rho_{\sigma}(\omega, p)$ at the same temperature and chemical potential.
In (a), the positions of the thresholds for $2\sigma$ and $2\pi$ decay channel determined by Eq. (\ref{threshold})
are denoted by \textcircled{s}, \textcircled{p}.}
\label{SpectP50_1}
\end{figure}
Figure \ref{SpectP50_1} is the result of
$\rho_{\sigma}(\omega, p)$ at $(T, \mu)=(5.1\MeV, 286.3 \MeV)$.
There is a sharp peak at $\omega=250 \MeV$ and a relatively small bump in the space-like region $\omega < \pab$
in Fig. \ref{SpectP50_1}(a): They correspond to the sigma meson with a modified mass at finite
temperature and the phonon mode composed of particle-hole excitations,
respectively, which is in accord with the result in RPA in \cite{Fujii:2004jt}.
In Fig. \ref{SpectP50_1}(b), the dispersion relations of
the sigma meson mode and particle-hole mode can be seen clearly.

The spectral function also tells us the decay and absorption processes
of the particle excitations from the width of the corresponding  peaks or bumps.
In our energy scale, the following processes contribute to the spectral function $\rho_{\sigma}(\omega,\vec{p})$:
\[
\sigma^{\ast}\rightarrow \sigma\sigma, \quad \sigma^{\ast}\rightarrow \pi\pi
, \quad \sigma^{\ast}\rightarrow \pbar \psi, \quad
\sigma^{\ast}\sigma\rightarrow \sigma,\quad
\sigma^{\ast}\pi\rightarrow \pi, \quad
\sigma^{\ast}\psi \rightarrow \psi,
\]
where $\sigma^{\ast}$ denotes a virtual state in the sigma channel
with energy-momentum $(\omega, \vec{p})$. The energy-momentum
conservation gives constraints on the possible $(\omega, \, \vec{p})$
region for the former three processes as follows:
\begin{align}
\omega \geq \sqrt{\vec{p}^{2}+(2M_{\sigma})^{2}}\ \ &\text{for}\ \
\sigma^{\ast}\rightarrow \sigma\sigma, \notag \\
\omega \geq \sqrt{\vec{p}^{2}+(2M_{\pi})^{2}}\ \ &\text{for}\ \
\sigma^{\ast}\rightarrow \pi \pi, \label{threshold}\\
\omega \geq \sqrt{\vec{p}^{2}+(2M_{\psi})^{2}}\ \ &\text{for}\ \
\sigma^{\ast}\rightarrow \pbar\psi, \notag
\end{align}
which are all in the time-like region. On the other hand, the latter
three processes are all collisional ones and possible only in the
space-like region, $0\leq \omega < \pab$.
In particular, the last process $\sigma^{\ast}\psi \rightarrow \psi$
corresponds to the absorption process of the $\sigma^{\ast}$ mode into a thermally excited quark.
In Fig. \ref{SpectP50_1}(a), the positions of the thresholds for $2\sigma$ and $2\pi$ decay channel determined by Eq. (\ref{threshold})
are shown, and the bumps corresponding to these processes
can be seen.

\subsection{Spectral functions near the QCD CP}
\label{SSSpectral}
\begin{figure}[!t]
\centering\includegraphics[width=0.9\columnwidth]{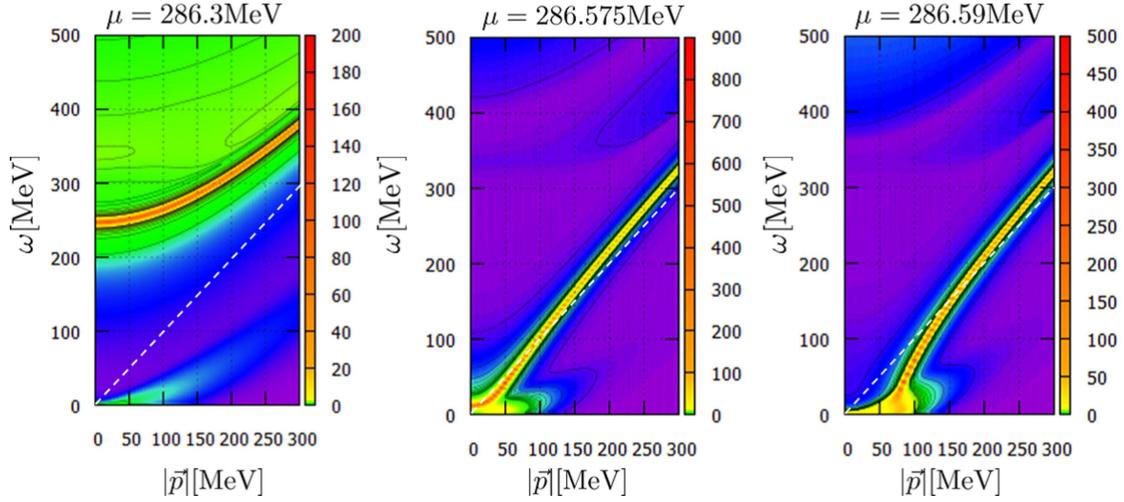}
\caption{Contour maps of $\rho_{\sigma}$
at $T=T_{c}$ and $\mu=286.3\MeV$, $286.575\MeV$ and
$286.59\MeV$.
(Taken from  \cite{Yokota:2016tip})}
\label{Spect3D}
\end{figure}

We calculate the  spectral function in the $\sigma$ channel near the QCD
CP, by increasing the chemical potential toward $\mu_{c}$
along constant temperature line $T=T_{c}$.
As seen in Sec. \ref{SSExample}, we can see the dispersion relations of the modes by making contour maps
of the spectral functions as functions of $\omega$ and $\pab$.
Figure~\ref{Spect3D} shows the dispersion relations of the sigma meson and
particle-hole modes near the CP.
At $\mu=286.3\MeV$, the sigma-mesonic peaks can be seen in the time-like
region as well as the particle-hole bump in the space-like region.
As the chemical potential increases, the dispersion relation
of the sigma-mesonic mode shifts downward and it touches  the light cone
near $\mu=286.575\MeV$. At $\mu=286.59\MeV$, in low-momentum region
the sigma-mesonic mode clearly penetrates into space-like region and
merges to the particle-hole bump, which has a  flat dispersion
relation in the small momentum region.

\begin{figure}[!t]
\centering\includegraphics[width=0.7\columnwidth]{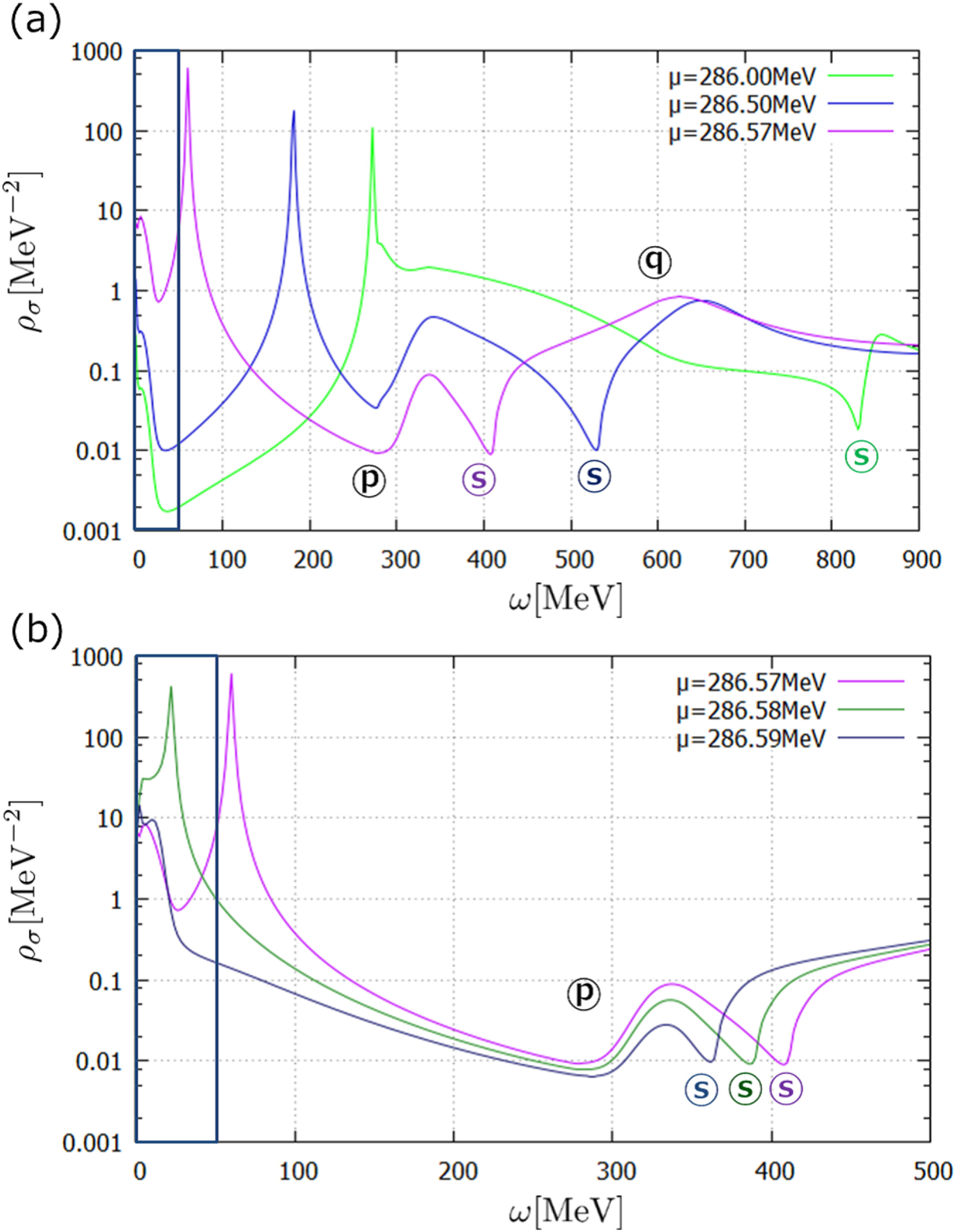}
\caption{$\rho_{\sigma}$ near the QCD CP at $T=T_{c}$:
(a) the results in $286.00\,\MeV \leq \mu \leq 286.57\,\MeV$ and (b) the results in $286.57\,\MeV \leq \mu \leq 286.59\,\MeV$.
The spatial momentum is set to $\pab=50\,\MeV$ and the inside of the blue box in each figure is the space-like region.
The $2\sigma$ decay thresholds for each chemical potential
are denoted by \textcircled{s}.
The $2\pi$ and $\pbar \psi$ decay thresholds hardly change
and are represented by \textcircled{p} and \textcircled{q}. (Taken from  \cite{Yokota:2016tip})
}
\label{SpectP50_2}
\end{figure}

Next, we show the strength of peaks and bumps of the spectral function when $p$ is set to $50 \MeV$.
The results at $\mu=286.00\MeV$, $\mu=286.50\MeV$ and
$\mu=286.57\MeV$ are shown in Fig. \ref{SpectP50_2} (a).
One can see the sigma-mesonic peak as well as bumps corresponding
to $2\sigma$ and $2\pi$ decay in the time-like region.
The peak position of the sigma-mesonic mode shifts to the lower energy
as the system approaches the CP.
The position of the $2\sigma$ threshold also shifts to a lower energy
while those of the $2\pi$ and $\pbar \psi$ thresholds hardly
change. The spectral function in the space-like region is drastically
enhanced as the system is close to the CP.
This behavior can be interpreted as the softening of the
particle-hole mode.
In Fig.~\ref{SpectP50_2}(b), we show the results at chemical potentials much
closer to the CP. Because of numerical instability
in $286.60\MeV \leq \mu \lesssim 360\MeV$,
we choose $\mu =286.58\MeV$ and
$\mu =286.59\MeV$.
For comparison, the result at $\mu=286.57\MeV$ is also shown.
These results are quite different from those in
$\mu \leq 286.57\MeV$. In $\mu >286.57\MeV$, the peak of the
sigma-mesonic mode penetrates into the space-like region and then
merges into the particle-hole mode.
Our results indicate that the sigma-mesonic mode as well as the
particle-hole mode can become soft near the CP.

\begin{figure}[!t]
\centering\includegraphics[width=0.7\columnwidth]{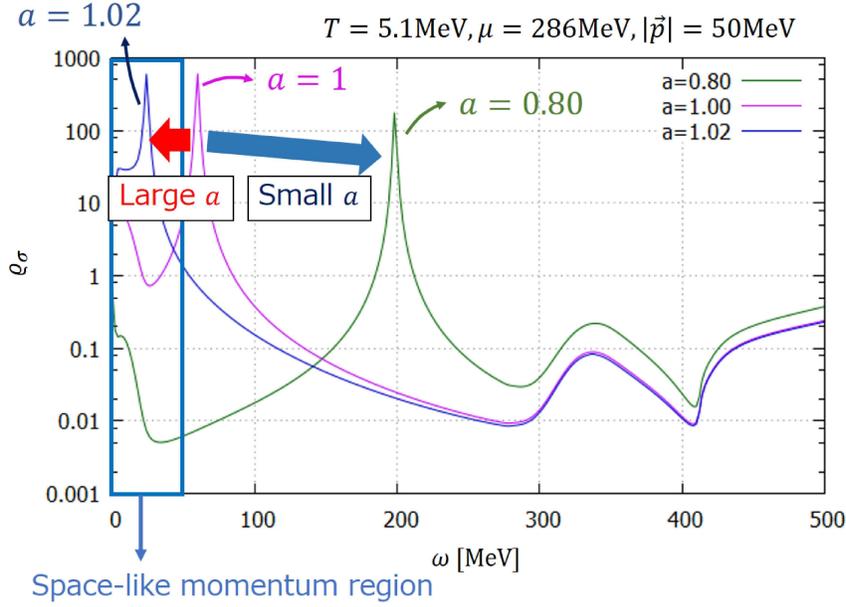}
\caption{$\rho_{\sigma}$ with a substituted three-point vertex
 $a\Gamma_{k,\sigma\sigma\sigma}^{(3)}$ at $T=T_{c}$,
 $\mu=286.57\,\mathrm{MeV}$, and $\pab=50\,\mathrm{MeV}$. The inside of the blue box is the space-like region.}
\label{G3change}
\end{figure}

One of the possible triggers of this phenomenon is the level
repulsion between the sigma-mesonic mode and other modes.
In particular, the two-sigma ($\sigma\sigma$) mode is considered to play an important role in
the level repulsion since the threshold of the two-sigma mode shifts
downward as the system approaches the CP.
Let us suppose that the particle--hole mode, the sigma-mesonic mode, and the two-sigma mode can each be described by a state having a single energy level.
Then the system can be regarded as a three-level system.
The interaction within the three states leads to a level repulsion:
If the interaction between the $\sigma$-mesonic mode and  the
$\sigma\sigma$ state becomes sufficiently strong as the system
approaches the CP,
the energy level of the sigma meson will be so strongly pushed down
that it penetrates into the space-like region.
To show that this scenario can be the case,
we change the strength of the three-point vertex
$\Gamma_{k,\sigma\sigma\sigma}^{(0,3)}$
by hand to investigate the behavior of the sigma-meson peak.
The results in the cases of multiplying
$\Gamma_{k,\sigma\sigma\sigma}^{(0,3)}$ by factors $0.8$ and $1.02$
are shown in Fig. \ref{G3change}. The position of the sigma meson
goes up when the three-point vertex is weakened, whereas it exhibits a
downward shift to a lower energy when the three-point vertex is slightly
enhanced. This result suggests that the above interpretation in terms of a level repulsion can be correct.

Here it should be noted that our results exhibit a superluminal group velocity of
the sigma-mesonic mode near the CP, as seen in
Fig.~\ref{Spect3D} for $\pab=100\,\MeV$ at $\mu=286.59\,\MeV$.
Such an unphysical extreme behavior may
be an artifact of our truncation scheme, in which some of the higher-order terms in
the derivative expansion
and the use of the three-dimensional regulator, Eqs.~\eqref{eq:reg_boson} and \eqref{eq:reg_fermion}
 \footnote{We thank J.Pawlowski for pointing out this possibility.},
although a drastic softening of the sigma-mesonic mode may persist.
We expect that such a drawback could disappear
if one
incorporates higher-derivative terms and/or uses a more sophisticated regulator respecting
the covariance as much as possible.
One of the most important
higher-derivative terms may be the wave-function renormalization
\cite{Helmboldt:2014iya,Kamikado:2013pya},
since the relative strengths of the modes
should be properly taken into account
 when collective modes are dynamically generated
 in addition to the modes described by the fields existing
in the bare Lagrangian, as is the case in the present work.

So far, we have concentrated on the spectral function in the sigma channel and
seen interesting behaviors of it near the CP. It would be intriguing
to examine whether the spectral function $\rho_{\pi}(\omega, p)$
in the pion channel shows any peculiar behavior near the CP.
In contrast to $\rho_{\sigma}$,
the dispersion relation of the pion mode stays in the time-like region and
$\rho_{\pi}$ hardly changes near the CP indicating that there is no critical behavior in the isovector pseudo-scalar
modes both in the space-like and time-like regions.

\section{Summary} \label{SSummary}
In this note, we have demonstrated how to compute spectral functions in
a relativistic system at finite temperature and density
within the functional renormalization group approach.
Our method is based on the local potential approximation (LPA)
and Litim's optimized regulator in three dimension, which enable
us to carry out an analytic continuation from imaginary to real
frequency at the level of the flow equations for the two-point
functions \cite{Tripolt:2013jra,Tripolt:2014wra}. We have also given a
detailed numerical procedure including a stability condition \cite{Yokota:2016tip}.

We have applied the method to the two-flavor quark-meson model which is composed
of light quarks, $\sigma$ and $\pi$ mesons as an effective realization of
spontaneous chiral symmetry breaking and restoration of QCD at low energies.
We have focused on the spectral
function of the scalar ($\sigma$) channel in the vicinity of the CP
which is located in large quark chemical potential and low temperature.
In this region, an explicit breaking of the chiral symmetry due to
a nonzero pion mass and the coupling of the quark density to the scalar
channel give a non-trivial structure to the spectral function. Indeed,
it was suggested that the particle-hole mode is enhanced near the critical
point and thus the density fluctuations are the soft modes at the
CP in a similar model calculation based on random phase
approximation~\cite{Fujii:2004jt}. We have shown that the particle-hole
mode has a growing support in the space-like region of the spectral
function as the system approaches to  the CP.
Furthermore,
we have found an anomalous dispersion relation of the $\sigma$ meson;
it penetrates into the space-like region as the system approaches to the
CP, and then merges into the particle-hole mode. This
anomalous softening of the $\sigma$ meson might be attributed to a
level repulsion between the $\sigma$ meson and the two-$\sigma$ mode.

Our results, obtained by calculating the spectral function of the
scalar channel with FRG, may imply a novel picture of the soft modes
of the QCD CP, which could influence the dynamical universality
class. Since our method is based on
LPA and the specific regulator function, the results should be examined
by improving the truncation scheme, e.g., including the wave-function
renormalization, and by exploring the regulator
dependence. Nevertheless, our result paves the way for investigating
emergent collective modes at finite temperature and density with the functional renormalization group.

\section*{Acknowledgments}
T.~Y.~ was supported by the Grant-in-Aid for JSPS Fellows (No. 16J08574).
T.~K.~ was partially supported by a Grant-in-Aid for Scientific Research from the Ministry of Education, Culture, Sports, Science and Technology (MEXT)
of Japan (Nos.16K05350,15H03663), by the Yukawa International
Program for Quark-Hadron Sciences.
K.~M.~ was supported by the Grantsin-Aid
for Scientific Research on Innovative Area from
MEXT (No. 24105008) and Grants-in-Aid for Scientific
Research from JSPS (No. 16K05349).
Numerical computation in this work was carried out at the
Yukawa Institute Computer Facility.

%%%%%%%%%%%%%%%%%%%%%%%%%%%%%%%

\end{document}